\documentclass{emulateapj}
\usepackage{apjfonts}
\usepackage{epsf}
\newcommand{\Msun}{\mathrm{M}_{\odot}}
\newcommand{\OIII}{[\ion{O}{3}]}

\begin{document}

\slugcomment{Submitted to ApJ Letters}
\shortauthors{}
\shorttitle{}

\title{Shrinking the Braneworld: Black Hole in a Globular Cluster}

\author{ Oleg Y. Gnedin\altaffilmark{1},
         Thomas J. Maccarone\altaffilmark{2},
         Dimitrios Psaltis\altaffilmark{3},
         Stephen E. Zepf\altaffilmark{4}}

\altaffiltext{1}{Department of Astronomy, University of Michigan, 
   Ann Arbor, MI 48109; \mbox{\tt ognedin@umich.edu}}
\altaffiltext{2}{School of Physics and Astronomy, University of Southhampton,
   Southhampton, SO17 1BJ, UK; \mbox{\tt tjm@astro.soton.ac.uk}}
\altaffiltext{3}{Departments of Astronomy and Physics, University of Arizona, 
   Tucson, AZ 85721; \mbox{\tt dpsaltis@email.arizona.edu}}
\altaffiltext{4}{Department of Physics and Astronomy, Michigan State University,
   East Lansing, MI 48824; \mbox{\tt zepf@pa.msu.edu}}

\date{\today}

\begin{abstract}
Large extra dimensions have been proposed as a possible solution to
the hierarchy problem in physics.  One of the suggested models, the
RS2 braneworld model, makes a prediction that black holes 
evaporate by Hawking radiation on a short timescale that depends on
the black hole mass and on the asymptotic radius of curvature of the
extra dimensions.  Thus the size of the extra dimensions can be
constrained by astrophysical observations.  Here we point out that the
black hole, recently discovered in a globular cluster in galaxy NGC
4472, places the strongest constraint on the maximum size of the extra
dimensions, $L \lesssim 0.003$ mm.  This black hole has the virtues of
old age and relatively small mass.  The derived upper limit is within
an order of magnitude of the absolute limit afforded by
astrophysical observations of black holes.
\end{abstract}


\section{Introduction}
 \label{sec:intro}

A universe with large extra dimensions has been proposed as a possible
solution to the hierarchy problem in physics
\citep{arkani-hamed_etal98, randall_sundrum99}.  The maximum size of
the extra dimensions can be constrained, in general, by verifying
Newton's law of gravity on sub-mm scales in torsion-balance laboratory
experiments.  Current constrains are as small as $L \lesssim 0.044$~mm
\citep{kapner_etal07, geraci_etal08}.

For braneworld models with large extra dimensions, the maximum size
can also be constrained by a variety of astrophysical tests (see
\citealt{arkani-hamed_etal98} for a discussion).  In particular, in
the RS2 braneworld model, black holes are predicted to evaporate on
the short timescale (\citealt{emparan_etal03} and references therein)
\begin{equation}
 \tau_{\rm ev} \approx 120 \left({M \over \Msun}\right)^3 
              \left({L \over 1 \, \mathrm{mm}}\right)^{-2} \mathrm{yr}\;,
 \label{eq:tau}
\end{equation}
where $M$ is the black hole mass and $L$ is the asymptotic radius of
curvature of the extra dimensions. This theoretical prediction has
been derived using scaling arguments and the AdS/CFT correspondence,
and, therefore, does not suffer from typical astrophysical
complications (see, however, \citealt{fitzpatrick_etal06} for a
discussion).

\citet{psaltis07} applied this argument to the black hole in the
binary system XTE~J1118+480 in the Galaxy.  The mass of the black hole
has been dynamically measured to be $M = 8.5 \pm 0.6 \, \Msun$, and a
lower limit on the age derived from its kinematics and position above
the Galactic plane was found to be $\tau_{\rm age} \gtrsim 1.8\times 10^7$ yr.
Equation (\ref{eq:tau}) and the analysis of measurement uncertainties
then constrain the maximum size $L \lesssim 0.08$~mm.
\citet{johannsen_etal09} further showed that a rapid evaporation of a
black hole in a binary system would lead to an orbital period
derivative that is potentially measurable. They then used the observed
upper bound on the orbital period derivative of the binary that
harbors the black hole A0620$-$00 to reach a similar upper bound on
the asymptotic radius of curvature. The same technique was later
applied to the binary that harbors the black hole XTE~J1118$+$480 by
\citet{johannsen09}.

In this {\it Letter} we discuss how the recent discovery of a black hole in
an old globular cluster leads to a much stronger upper bound on the
asymptotic radius of curvature of the extra dimensions.

\section{An old Black Hole in a Globular Cluster}
 \label{sec:bh}

\citet{maccarone_etal07} reported the discovery of a compact object in
a globular cluster RZ2109 located in an elliptical galaxy NGC~4472
in the Virgo cluster of galaxies.  The black hole nature of this
object was inferred from the strong variability of the high-luminosity
($4\times 10^{39}$ erg s$^{-1}$) X-ray source observed with the
XMM-Newton satellite.  \citet{zepf_etal08} furthermore showed that the
optical spectrum is dominated by a very broad \OIII\ 5007\AA\ emission
line.  They argued that the only extant model able to account for the
observations is a $\simeq 10 \ \Msun$ black hole accreting close to
its Eddington limit.  We discuss further constraints on the mass in
\S\ref{sec:mass}.

The most interesting feature of this observation is that it is the
first robust identification of a stellar-mass black hole in a globular
cluster.  The association with a globular cluster strongly argues that
the black hole is very old.  Globular clusters are special stellar
systems, which remain gravitationally bound for a Hubble time and
contain stars that all formed within a relatively short period of
time.  Even in the cases of a few of the most massive clusters in the
Galaxy, for which detailed color-magnitude diagrams and spectroscopy
show evidence of multiple stellar populations, the likely age spread
is of the order of 1 Gyr \citep[e.g.,][]{sollima_etal05,
villanova_etal07, cassisi_etal08}.  This spread is still an order of
magnitude shorter than the cluster age.  Based on observations of the
main sequence turnoff point of the cluster stars, the oldest globular
clusters in the Galaxy are thought to be approximately 13 Gyr, and
none is younger than 7 Gyr \citep[e.g.,][]{krauss_chaboyer03,
sarajedini_etal07, marin-franch_etal09}.  Age measurements of
extragalactic clusters are more uncertain, but both absorption-line
spectroscopy \citep{cohen_etal03} and optical and near-infrared
photometry \citep{hempel_etal07} studies find that the globular
clusters in NGC 4472 are as old as those in the Galaxy.  Moreover, an
ongoing detailed study of the spectrum of RZ2109 itself confirms its
old age (Steele et al. in prep).  Thus a reliable and conservative age
estimate of the globular cluster RZ2109 is $10 \pm 3$ Gyr.

If the black hole is found to be in the globular cluster in the
present epoch, its progenitor must have formed with the rest of the
stars in the cluster.  This is because a globular cluster in a galaxy
cannot capture any star that was previously unbound to it
\citep[e.g.,][]{mieske_baumgardt07}.  The cluster gravitational
potential is too small compared to the kinetic energy of objects
orbiting a large elliptical galaxy such as NGC~4472.  A typical
velocity dispersion inside globular clusters is $\sim 10$ km s$^{-1}$,
while orbital velocities in elliptical galaxies are in excess of 200
km s$^{-1}$.

The lifetime of a progenitor star that led to the formation of the
black hole in a type II supernova explosion is negligibly short.
Stellar evolution models for a wide range of progenitor masses
indicate main-sequence lifetimes smaller than a few times $10^7$ yr.
As a result, the age of a $10\, \Msun$ black hole in a globular
cluster is practically equal to the age of the cluster itself.

\begin{figure}
\vspace{-0.4cm}
\centerline{\epsfxsize3.6truein \epsffile{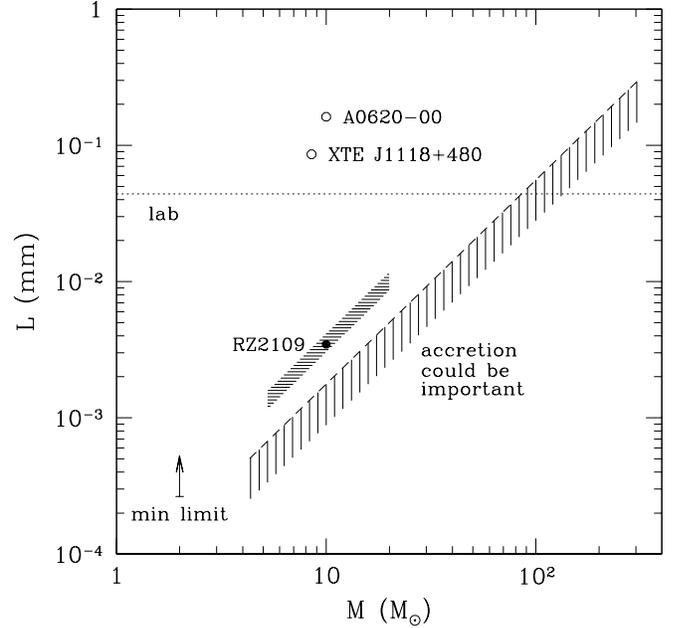}}
\vspace{-0.3cm}
\caption{Constraints on the asymptotic radius of curvature of the
  extra dimensions, $L$, from the evaporation of black holes of mass
  $M$ and age $\tau$ ($L \propto M^{3/2} \tau^{-1/2}$).  Horizontal
  dotted line shows the current laboratory limit.  Filled circle shows
  the limit obtained in this paper for the black hole in a globular
  cluster RZ2109.  Shaded region indicates the uncertainty range due to
  the expected uncertainty of black hole mass (from 5 to 20 $\Msun$)
  and its age (from 7 to 13~Gyr).  The minimum limit is obtained for
  the smallest black hole mass ($2\, \Msun$) and largest age
  (13.7 Gyr).  Dashed line with a shaded region extending downward shows
  the range of masses where steady accretion of matter onto the black
  hole may dominate the evaporation.}
\vspace{0.3cm}
  \label{fig:l}
\end{figure}

Using equation (\ref{eq:tau}) for the age of 10 Gyr and mass of $10 \
\Msun$, we obtain an upper bound on the asymptotic radius of curvature
of the extra dimensions $L \lesssim 0.003$~mm.  This bound is lower by
an order of magnitude than previous table-top and astrophysical
constraints.  Figure~\ref{fig:l} and Table~\ref{tab:obs} summarize the
constraints on the maximum size of the extra dimensions from the black
hole in RZ2109, the black holes in XTE J1118+480 and A0620$-$00, and
the laboratory measurements.  The new limit provided by the black hole
in RZ2109 is much smaller than that for XTE J1118+480 because of the
crucial difference in the black hole ages.  The shaded region shows
the expected uncertainty in this limit due to the uncertainties of the
black hole mass (between 5 and 20 $\Msun$) and its age (between 7 and
13~Gyr).  Even for the worst combination of the parameters, the upper
bound increases to $L\lesssim 0.01$~mm, which is still an improvement
by a factor of $\simeq 5$ from the current laboratory limit.

\begin{table}
\begin{center}
\caption{\sc Observations of Black Holes}
\label{tab:obs}
\begin{tabular}{lccl}
\tableline\tableline\\
\multicolumn{1}{l}{BH system} &
\multicolumn{1}{c}{$M$ ($M_{\sun})$} &
\multicolumn{1}{c}{age (yr)} &
\multicolumn{1}{c}{$L$ (mm)}
\\[2mm] \tableline\\
XTE J1118+480 & $8.5 \pm 0.6$ & $>10^7$    & 0.08 \\
A0620-00      & $10 \pm 5$    & ...        & 0.16 \\
RZ2109        & $\approx 10$  & $\approx 10^{10}$ & 0.003 \\
\tableline
\end{tabular}
\end{center}
\vspace{0.4cm}
\end{table}

\section{Caveats and Uncertainties}
 \label{sec:discussion}

\subsection{Accretion and Collisions}

In this {\em Letter} we placed a strong upper bound on the asymptotic
radius of curvature of the extra dimensions, by estimating the age and
mass of the black hole recently discovered in an extragalactic
globular cluster.  The only potential caveats to our arguments would
appear if the black hole were less massive when it formed and then
subsequently merged with other stars or has grown significantly by
accretion.  While such scenarios are possible in a globular cluster,
we argue here that they are unlikely to change our quantitative bound
on the size of the extra dimensions.

The black hole mass can change with time because of evaporation,
accretion, and collisions with other stars:
\begin{equation}
 {dM \over dt} = -{M \over 3 \tau_{\rm ev}(M)} 
    + \dot{M}_{\rm acc} + \dot{M}_{\rm coll}.
 \label{eq:coll}
\end{equation}
The rates of accretion and collisions are likely to be variable in
time, but to obtain practical estimates we make the following
simplifying assumptions.  We take $\dot{M}_{\rm coll}$ to be constant
in time and independent of the BH mass.  We take the accretion rate to
be a constant fraction $f_E \le 1$ of the Eddington rate,
\begin{equation}
  \dot{M}_{\rm acc} = f_E {M \over t_E}, \quad 
  t_E \approx 4.5\times 10^7 \left({\epsilon \over 0.1}\right)\ {\rm yr},
\end{equation}
where $\epsilon$ is the radiative efficiency of accretion and is
typically $\epsilon \approx 0.1$.  Thus $\dot{M}_{\rm acc} \propto M$.
With these assumptions the total mass derivative is a monotonic
function of the mass:
\begin{equation}
  {dM \over dt} = -{\, \rm const}{L^2 \over M^2}
    + {f_E \over t_E} M + \dot{M}_{\rm coll},
\end{equation}
that is $dM/dt$ increases if $M$ increases, and vice versa.  So if at
one time in the past the derivative was positive the mass will only
keep increasing, the rate of evaporation will diminish, and the black
hole will survive.  Conversely, if the derivative was negative both
accretion and collisions will become less important with time, and the
black hole will evaporate in a finite amount time similar to
$\tau_{\rm ev}(M)$.

Consider first the case of continuous accretion at a fraction $f_E$ of
the Eddington limit.  The increase of mass by accretion would
overwhelm the evaporation of the black hole if its mass is larger than
\citep{psaltis07}
\begin{equation}
  M \ge 50 \left({1\over f_E}\right)^{1/3} \left(\frac{\epsilon}{0.1}\right)^{1/3}
   \left(\frac{L}{1~\mbox{mm}}\right)^{2/3} \ \Msun\,.
  \label{eq:acc}
\end{equation}
Stellar black holes have a minimum mass $\ga 2\ \Msun$, which is an
upper limit for a neutron star mass for most proposed equations of
state of nuclear matter.  We can set an upper bound on the
time-averaged value of $f_E$ by requiring that the black hole has
increased its mass at most by a factor of $n \la 10$ over its age,
$\tau_{\rm age}$.  This leads to
\begin{equation}
  f_E \la {t_E \over \tau_{\rm age}} \ln{n} = 4.5\times 10^{-3} \; \ln{n}
    \left(\frac{\epsilon}{0.1}\right)
    \left(\frac{\tau_{\rm age}}{10~\mbox{Gyr}}\right)^{-1}.
\end{equation}
As a result, the black hole could have accreted only at a fraction of
a percent of the Eddington rate over its lifetime.  An alternative
interpretation is that the black hole could have accreted at the
Eddington rate but only for a fraction $f_E$ of its lifetime.  In any
case, if the observed black hole mass has been gained over time, the
condition of equation (\ref{eq:acc}) must also be satisfied at some
earlier time when the mass was lower.  In our example the initial mass
was $M_{\rm obs}/n$ and hence accretion dominates evaporation at
all times if the current mass is higher than
\begin{equation}
  M \ga 300 \; {n \over (\ln{n})^{1/3}} 
    \left(\frac{\tau_{\rm age}}{10~\mbox{Gyr}}\right)^{1/3}
    \left(\frac{L}{1~\mbox{mm}}\right)^{2/3} \ \Msun\,.
  \label{eq:macc}
\end{equation}
The more mass a black hole gained by accretion the stricter is this
limit.  Figure \ref{fig:l} shows the limit for a black hole that
doubled its mass over its lifetime ($n=2$).  This limit comes within a factor
of 2 in mass $M$ or size $L$ obtained for RZ2109.  Although
interestingly close, this limit indicates that steady accretion of
matter could not have significantly hampered the evaporation of the
black hole.

Consider now the case in which the mass of the black hole has grown
after it merged with other stars in the cluster.  The increase in mass
by a factor of $n$ requires a time-averaged rate of $\dot{M}_{\rm
coll} = M \tau_{\rm age}^{-1} (1-1/n)$.
This rate dominates the evaporation rate for
\begin{equation}
  M \ga 300 \; {n \over (n-1)^{1/3}} 
    \left(\frac{\tau_{\rm age}}{10~\mbox{Gyr}}\right)^{1/3}
    \left(\frac{L}{1~\mbox{mm}}\right)^{2/3} \ \Msun\,.
\end{equation}
This equation is essentially the same as equation (\ref{eq:macc}),
except for a factor $(n-1)^{1/3}$ instead of $(\ln{n})^{1/3}$.  As in
the case of accretion, the strictest constraint on the mass occurs at
the initial epoch when the black hole mass is the smallest.
Therefore, collisions can be important only in the same mass range as
steady accretion.

\subsection{Mass of the black hole in RZ2109}
  \label{sec:mass}

The constraint on $L$ scales with the mass of the black hole in
RZ2109.  How reliable is our estimate of its mass of $10\ \Msun$?
The strong \OIII\ emission from RZ2109 strengthens the case for a
stellar mass black hole in the cluster in three key ways.

First, it dramatically reduces the already low probability that the
variable X-ray source in the cluster is a background active galactic
nucleus superposed on the globular cluster in the galaxy, by providing
strong evidence that there is an unusual feature at the same redshift
as the cluster.

Second, the \OIII\ lines provide evidence against an alternative
possibility that the source is an intermediate mass black hole
accreting well below its Eddington limit.  The luminosity of the
\OIII\ 5007\AA\ line is about $1.4 \times 10^{37}$ erg s$^{-1}$
\citep{zepf_etal08}, while the velocity width of the line is about
2000 km s$^{-1}$.  This combination cannot be produced by virial
motions around a black hole of less than about $10^4 \ \Msun$ without
invoking a density of oxygen atoms exceeding the critical density for
the \OIII\ line \citep{zepf_etal08}, and even at $10^4 \ \Msun$,
considerable fine tuning is needed to have the full volume of the
"virial region" at exactly the critical density of the \OIII\ lines.
The most likely situation, then, is that the emission line's large
velocity width comes from a strong wind.  While energetically
important disk winds have been suggested from sources accreting at low
fractions of the Eddington limit
\citep[e.g.,][]{blandford_begelman99}, only at very high fractions of
$L_E$ are winds expected with the mass loss rates needed to produce
the observed \OIII\ emission \citep[e.g.,][and references
therein]{proga07}.  Therefore, the strength and breadth of the \OIII\
line indicates that the accretor is an object of stellar mass
accreting at or slightly above its Eddington luminosity.


Third, the bright \OIII\ emission presents a strong case that the
source is not a neutron star accreting well above its Eddington
luminosity.  In \citet{maccarone_etal07} it was shown that, without
beaming, an unphysically high accretion rate would be required to
allow for this system to be a super-Eddington neutron star accretor.
However, geometric beaming has been suggested as a way to produce the
high luminosities seen in some ultraluminous X-ray sources without
invoking unphysically high mass transfer rates and without invoking
black hole masses in excess of $\sim 20 \ \Msun$
\citep[e.g.,][]{king_etal01}.  The observed \OIII\ emission is a new
argument against significant beaming.  The \OIII\ emission cannot come
from a small enough region to be beamed geometrically, and hence must
be unbeamed.  For the beaming factors of $\sim 10$, which would be
needed to allow the accretor in RZ2109 to be a neutron star, one would
first need to explain an intrinsic luminosity ratio $L_{\rm [O\,
III]}/L_{\rm bol} \sim 0.03$, far below what is seen from active
galactic nuclei \citep[e.g.,][]{heckman_etal05}.  Also, one would need
to explain the lack of any globular clusters with strong, broad \OIII\
lines but no bright X-ray sources, given that a large number of
globular clusters have been searched for emission lines,
turning up no other cases of clusters with very broad lines but
finding numerous narrow lines typically associated with planetary
nebulae \citep[e.g.,][]{minniti_rejkuba02, brodie_etal05,
pierce_etal06, chomiuk_etal08}.

Thus we conclude that the accretor in RZ2109 is a black hole of
$5.3-20 \ \Msun$.  These limits come from estimating the fraction of
the Eddington luminosity that is contributed by the observed X-ray
luminosity $L_X \approx 4\times 10^{39}$ erg s$^{-1}$, as follows:
\begin{equation}
  M \approx 16 \left({L_E \over L_X}\right) \left({2\over \mu_e}\right) \ \Msun,
\end{equation}
where $\mu_e$ is the mean molecular weight per electron.  The lack of
Balmer emission from the source indicates that the donor star is
likely to be hydrogen poor, which gives $\mu_e \approx 2$.  Such a
donor could be a white dwarf in a $\sim 5$ minute orbit, consistent
with persistent emission.  The lack of emission from any element
heavier than oxygen is also a strong argument against the possibility
of this system being a tidally-induced supernova
\citep[suggested by][]{rosswog_etal09}.

The lower bound on the mass can be estimated by looking at the degree
to which black hole X-ray binaries in the Galaxy can exceed their
Eddington luminosities in the absence of beaming effects.  In the
compilation of \citet{garcia_etal03} the only black hole X-ray binary
observed at more than 3 times its Eddington luminosity is V4641 Sgr,
which shows highly superluminal proper motions of its jet
\citep{orosz_etal01} and hence is likely to be strongly beamed, and
which also showed such a high luminosity only for a very short
duration, in contrast to the steady high $L_X$ of the source in
RZ2109.  Thus we can place a lower bound on the black hole mass by
taking $L_X \le 3 \, L_E$, which gives $M \ge 5.3 \ \Msun$.

The upper bound comes from the conservative assumption that the source
must be at least at 80\% of its Eddington luminosity to be
driving a strong wind.  Taking $L_X \ge 0.8 \, L_E$ gives $M \le
20 \ \Msun$, in the case of accretion of hydrogen-free material.  Note
that if the accreted material is all hydrogen, then $\mu_e=1$ and the
upper bound increases to $40 \ \Msun$, but the observed lack of Balmer
emission makes the latter case unlikely.

\section{Summary}

We set the strongest upper limit to date on the asymptotic radius of
curvature of the extra dimensions, $L \lesssim 0.003$ mm, based on
observations of the black hole in a globular cluster RZ2109 in an
external galaxy NGC 4472.  This limit scales with the age and mass of
the black hole, but for all realistic values of these parameters, the
robust bound is $L < 0.01$ mm (Fig. \ref{fig:l}).

The bound can be further reduced if a smaller mass for an old black
hole is reliably measured.  The absolute minimum is obtained for the
smallest black hole mass and oldest age.  The smallest stellar black
hole mass is expected to be around $2\, \Msun$.  The oldest age cannot
exceed the age of the universe, 13.7 Gyr.  Thus the strongest
constraint on $L$ from black hole evaporation is $\sim 3\times
10^{-4}$ mm.  It is shown by the upward arrow on Figure~\ref{fig:l}.
The current constraint derived for the black hole in RZ2109 is within
an order of magnitude of this absolute limit afforded by astrophysical
observations of black holes.

\acknowledgements 

We thank Jon Miller and Tim Johannsen for discussions.  SZ is
supported in part by NSF grant AST-0807557.  DP is supported by NSF
CAREER award NSF~0746549.  OG is supported in part by NSF grant
AST-0708087.

\bibliography{extrad}

\end{document}